\documentclass[aps,prd,reprint,superscriptaddress,preprintnumbers,showpacs,nofootinbib]{revtex4-1}

\usepackage[colorlinks, pdfborder={0 0 0}, plainpages=false]{hyperref}
\usepackage[utf8]{inputenc}
\usepackage{graphicx}
\usepackage{amsmath}
\usepackage{amssymb}
\usepackage{acronym}
\usepackage{amsfonts,pifont}
\usepackage{txfonts}
\usepackage[normalem]{ulem}
\usepackage{booktabs}
\usepackage{multirow}

\newcommand{\cardiff}{School of Physics and Astronomy, Cardiff University, The
Parade, Cardiff, CF24 3AA, United Kingdom}

\newcommand{\X}{$\mathbf{X}\,$}
\newcommand{\Y}{$\mathbf{Y}\,$}
\newcommand{\floor}[1]{\left\lfloor #1 \right\rfloor}
\newcommand{\expnumber}[2]{{#1}\mathrm{e}{#2}}

\newacro{BH}{Black Hole}
\newacro{BBH}{Binary Black Hole}
\newacro{NR}{Numerical Relativity}
\newacro{GW}{Gravitational Wave}
\newacro{IMR}{Inspiral-Merger-Ringdown}
\newacro{NS}{Neutron Star}
\newacro{BNS}{Binary Neutron Star}
\newacro{LSC}{LIGO Scientific Collaboration}
\newacro{LVC}{LIGO-Virgo Collaboration}
\newacro{LIGO}{Laser Interferometric Gravitational-Wave Observatory}
\newacro{aLIGO}{Advanced LIGO}
\newacro{Adv}{Advanced Virgo}
\newacro{EM}{electromagnetic}
\newacro{PN}{Post Newtonian}
\newacro{EOB}{Effective-One-Body}
\newacro{MSA}{multiple scale analysis}
\newacro{SUA}{shifted uniform asymptotics}
\newacro{SPA}{stationary phase approximation}
\newacro{SNR}{signal-to-noise ratio}
\newacro{PSD}{power spectral density}
\newacro{ROQ}{reduced order quadrature}

\begin{document}

\title{Gravitational-wave surrogate models powered by artificial neural networks:\protect\\The ANN-Sur for waveform generation}

\author{Sebastian Khan}
\affiliation\cardiff

\author{Rhys Green}
\affiliation\cardiff


\date{\today}

\begin{abstract}

Inferring the properties of black holes and neutron stars is a key science goal
of gravitational-wave (GW) astronomy. To extract as much information as possible
from GW observations we must develop methods to reduce the cost of Bayesian
inference. In this paper, we use artificial neural networks (ANNs) and the
parallelisation power of graphics processing units (GPUs) to improve the
surrogate modelling method, which can produce accelerated versions of existing
models. As a first application of our method, ANN-Sur, we build a time-domain
surrogate model of the spin-aligned binary black hole (BBH) waveform model
SEOBNRv4. We achieve median mismatches of $\sim \expnumber{2}{-5}$ and
mismatches no worse than $\sim \expnumber{2}{-3}$. For a typical BBH waveform
generated from $12 \,\rm{Hz}$ with a total mass of $60 M_\odot$ the original
SEOBNRv4 model takes $1812 \, {\rm{ms}}$. Existing bespoke code optimisations
(SEOBNRv4opt) reduced this to $91.6 \, {\rm{ms}}$ and the interpolation based,
frequency-domain surrogate SEOBNRv4ROM can generate this waveform in $6.9 \,
{\rm{ms}}$. Our ANN-Sur model, when run on a CPU takes $2.7 \, {\rm{ms}}$ and
just $0.4 \, {\rm{ms}}$ when run on a GPU. ANN-Sur can also generate large
batches of waveforms simultaneously. We find that batches of up to $10^{4}$
waveforms can be evaluated on a GPU in just $163 \, {\rm{ms}}$, corresponding to
a time per waveform of $0.016 \, {\rm{ms}}$. This method is a promising way to
utilise the parallelisation power of GPUs to drastically increase the
computational efficiency of Bayesian parameter estimation.

\end{abstract}

\maketitle

\section{Introduction}
\label{sec:intro}

The swift and accurate computation of the gravitational-wave (GW) signal from
merging compact binaries is a crucial part of GW astronomy. Over the past few
years enormous progress has been made in modelling the GW signal
~\cite{Mehta2017AccurateIG,Mehta2019IncludingMM,Khan2020IncludingHO,
Estelles2020IMRPhenomTPAP,2018PhRvD..98h4028C,Nagar2018TimedomainEG,
Nagar2019AME,Rifat2020SurrogateMF,PhysRevD.99.064045,Khan2019PhenomenologicalMF,
PhysRevResearch.1.033015,Williams2019APN,Ossokine2020MultipolarEW,
Dietrich2019MatterII,Dietrich2019ImprovingTN,Pratten2020SettingTC,
London2018FirstHM,Nagar2020AME,Pratten2020LetsTA,GarciaQuiros2020IMRPhenomXHMAM}
and recent models have played imporant roles in the analysis of recent GW
events~\cite{LIGOScientific:2020stg,Abbott:2020khf}. However, as waveform models
relax simplifying approximations (such as including sub-dominant multipoles) the
computational cost tends to increase, which ultimately limits their use in GW
analyses.


To reduce the computational cost of generating waveforms the community has
developed several bespoke optimizations
~\cite{Devine2016OptimizingST,Knowles:2018hqq,PhysRevD.99.021501}. But these
typically require expert knowledge and might not provide general optimizations
that other models can incorporate. There are many methods to accelerate Bayesian
parameter estimation~\cite{PhysRevD.94.044031,
Morisaki2020RapidPE,Canizares2015AcceleratedGW,
Canizares2013GravitationalWP,Zackay2018RelativeBA,
Cornish2020BlackHH,Vinciguerra2017AcceleratingGW,
GarciaQuiros2020AcceleratingTE,Smith2014RapidlyET} but in general they each make
simplifying assumptions that mean not all waveform models can readily take
advantage of the potential speed-up. Another way to accelerate analyses is by
parallelisation. Typically this means parallelising your analysis across
multiple CPUs however, there has been growing interest in the use of graphics
processing units (GPUs),
see~\cite{Guo2017AccelerationOL,Keitel2018FasterSF,Katz2020GPUacceleratedMB,Talbot2019ParallelizedIF,Wysocki2019AcceleratingPI,Usman2015ThePS}
for applications in GW astronomy.

Alternatively, data-driven methods can be employed that are waveform model
agnostic and hence are of great interest.
One such method is called surrogate modelling~\cite{PhysRevX.4.031006}. Here,
one attempts to build a fast and accurate approximation (a surrogate or
emulator) of a slower model. A successful way to build these models typically
begins with building a reduced basis representation (e.g. a singular value
decomposition or greedy reduced basis) of the model
~\cite{PhysRevX.4.031006,Smith2013TowardsRP,
2014CQGra..31s5010P,Cannon2012InterpolatingCB,
Barta2018FastPA,Setyawati2019EnhancingGW,Cannon2013InterpolationIW}. One of the
biggest issues in reduced basis surrogate modelling is the approximation of the
reduced basis coefficients. This is a multidimension interpolation or regression
problem and has recently been investigated in~\cite{Setyawati:2019xzw} where the
authors systematically compared different interpolation and regression methods.

In this work we train artificial neural networks (ANNs), developed with the {\tt
TensorFlow}~\cite{tensorflow2015-whitepaper} library, to accurately and
efficiently estimate the projection coefficients of a reduced basis. ANNs are a
versitile tool~\cite{doi:10.1063/1.1144830} and have recently been applied to
solve reduced order modelling problems across multiple disciplines using a
non-intrusive
framework~\cite{Hesthavena2017NonintrusiveRO,Jacquier2021NonIntrusiveRM,2020PhyD..41232614G,MuckeNikolajTakata2019ROMf,SAN2019271}.
The use of ANNs in GW astronomy has increased recently
~\cite{Setyawati:2019xzw,Wong2020GravitationalWP,Wong2020GravitationalwaveSI,Gerosa2020GravitationalwaveSE,Marulanda2020DeepLG,2020arXiv200601509S,PhysRevD.100.044009,PhysRevD.101.064009,Gabbard2019BayesianPE,Haegel2020PredictingTP,Green2020GravitationalwavePE,Cuoco2020EnhancingGS,Lin2019BinaryNS,Shen2019DeepLA,Shen2019DenoisingGW,Carrillo2015TimeSA,2020arXiv200704176L,Graff2013SKYNETAE,2020arXiv200806071C}
and in particular~\cite{PhysRevLett.122.211101} where the authors used ANNs to
model the greedy reduced basis coefficients for a frequency domain inspiral
post-Newtonian waveforms in the context of massive binary black holes (BBHs)
that the space based GW observatory LISA~\cite{AmaroSeoane2017LaserIS} will be
sensitive to. Here we look at the projection coefficients of an empirical
interpolation basis for time domain waveforms. We generate the complete
inspiral, merger and ringdown waveform for the dominant ($\ell = |m| = 2$)
multipole of spin-aligned BBH coalescences using the SEOBNRv4
model~\cite{2017PhRvD..95d4028B}.

One advantage of our approach is that our ANN powered surrogate model (ANN-Sur)
can be executed on either a CPU or GPU because it is developed with {\tt
TensorFlow} and allows us to explore the possible benefits of utilising GPUs. We
find that by generating waveform on a GPU we gain a significant improvement in
computationally efficiency. On average, waveforms generated with ANN-Sur take
$2.7 {\rm{ms}}$ on a CPU, which corresponds to a speed-up factor of 661 when
compared to the SEOBNRv4 and a factor of 33 when compared to SEOBNRv4opt. Moving
waveform generation to a GPU provides a futher factor of 7 improvement taking
just $0.4 {\rm{ms}}$, which corresponds to a speed-up of 4646 (235) when
compared to SEOBNRv4 (opt). These improvements can be readily passed on to
standard parameter estimation codes.

Our model can also generate large batches of waveforms
simultaneously~\cite{Chua2020LearningBP}. We find that batches of waveforms up
to sizes of $10^4$ take only $163 {\rm{ms}}$ on the GPU,\footnote{We were
limited to batches $\mathcal{O}(10^4)$ due to GPU memory limitations.} a factor
of $\sim 30$ times faster than the CPU. We estimate that the time taken to
generate the same waveforms using the SEOBNRv4opt model, on a single CPU, would
take $\mathcal{O}(15) {\rm{mins}}$, corresponding to a speed-up factor of $\sim
5000$. These results are encouraging and suggest a way to drastically
drastically reduce waveform generation times using GPUs.

\section{Method}
\label{sec:method}

Let $h(t) = h_{+}(t) - i h_{\times}(t)$ be the predicted complex
gravitational-wave strain from a fiducial model, where $t$ is the time. We
expand this in terms of a spin-weight $-2$ spherical harmonic basis, which
allows us to separate out the intrinsic parameters $\boldsymbol{\lambda}$ (black
hole component masses and spin angular momenta) from extrinsic parameters
$(\theta,\varphi)$ (direction of propagation)

\begin{equation}
h(t;\boldsymbol{\lambda};\theta,\varphi) = \sum_{\ell \geqslant 2} \sum_{-\ell\leqslant m \leqslant \ell} h_{\ell,m}(t,\boldsymbol{\lambda}) {}_{-2}Y_{\ell,m}(\theta,\varphi) \, .
\end{equation}

If we restrict ourselves to non-eccentric binary black
hole systems with spins either alligned or anti-alligned with respect to the
orbital angular momentum then the system is completely specified by it's
mass-ratio $q = m_1/m_2$ ($m_1$ and $m_2$ are the primary and secondary masses
respectively), and the components of the individual BH spin vectors that are
aligned with the orbital angular momentum $(\chi_{1}, \chi_{2})$. Furthermore we
will model the $(\ell, m) = (2,\pm 2)$ multipoles which are the dominant
multipoles for comparable mass BBH systems. The method we use is agnostic to the
the specific GW multipole and can therefore be applied to the other multipoles
in a similar way, however, here we are interested in developing our method and
restrict outselves to just the dominant mulitpoles. As a final simplification we
note that for aligned-spin binaries the $(2,2)$ and $(2, -2)$ multipoles are
related to eachother according to $h_{2,2}(t) = h^{*}_{2, -2}(t)$, where $*$
denotes the complex conjugation. Therefore, we will only model the $h_{2,2}(t;
\boldsymbol{\lambda})$ data where $\boldsymbol{\lambda} = (q, \chi_{1},
\chi_{2})$. Instead of modelling the real and imaginary parts of
$h_{2,2}(t;\boldsymbol{\lambda})$ as done in~\cite{PhysRevLett.122.211101} we
decompose the data into an amplitude, $A(t;\boldsymbol{\lambda}) \equiv
|h_{2,2}(t;\boldsymbol{\lambda})|$, and phase, $\phi(t;\boldsymbol{\lambda})
\equiv \rm{arg}(h_{2,2}(t;\boldsymbol{\lambda}))$ and model these
independently~\cite{PhysRevX.4.031006}. The original complex data is recovered
with

\begin{equation}
h_{2,2}(t;\boldsymbol{\lambda}) = A(t;\boldsymbol{\lambda}) e^{-i
\phi(t;\boldsymbol{\lambda})} \, .
\end{equation}

We use the surrogate modelling methods described in~\cite{PhysRevX.4.031006,
PhysRevD.95.104023}, borrowing notation and only recounting the basic steps
here. We aim to build a surrogate model of the GW signal, denoted $h^{S}(t;
\boldsymbol{\lambda})$, that emulates the fiducial model such that $h^{S}(t;
\boldsymbol{\lambda}) \approx h(t; \boldsymbol{\lambda})$ to within a given
error tolerance. The surrogate model is defined for times $t \in [t_{\rm{min}},
t_{\rm{max}}]$ and for system parameters $\boldsymbol{\lambda} \in \mathcal{T}$,
where $\mathcal{T}$ is the compact parameter space of all possible BBH
parameters. We therefore aim to build computationally efficient and accurate
representations of the amplitude and phase functions
$A^{S}(t;\boldsymbol{\lambda})$ and $\phi^{S}(t;\boldsymbol{\lambda})$
respectively. With the final surrogate $h_{2,2}^{S}$ given by

\begin{equation}
h_{2,2}^{S}(t;\boldsymbol{\lambda}) = A^{s}(t;\boldsymbol{\lambda}) e^{-i
\phi^{S}(t;\boldsymbol{\lambda})} \, .
\end{equation}

In the following discussion we will use
$X(t;\boldsymbol{\lambda})$ as a placeholder variable to describe either the
amplitude or the phase and $X^{S}(t;\boldsymbol{\lambda})$ as the surrogate
approximation.

We can build an efficient representation of a function
$X(t;\boldsymbol{\lambda})$ by building a reduced basis. A reduced basis is a
linear decomposition such that for any value $\boldsymbol{\lambda} \in
\mathcal{T}$ we can approximate $X(t; \boldsymbol{\lambda})$ as
linear combination of projection coefficients $\{c_i(\boldsymbol{\lambda})\}_{i=1}^{n}$
and the $n$-element basis $B_n = \{e_i(t)\}_{i=1}^{n}$ given by

\begin{equation}
X(t; \boldsymbol{\lambda}) \approx \sum_{i=1}^{n} c_{i}(\boldsymbol{\lambda}) e_{i}(t) \, .
\end{equation}

We define the representation error between the true function and our reduced
basis approximation as $\sigma$

\begin{equation}
\sigma = \left\Vert  X(t; \boldsymbol{\lambda}) - \sum_{i=1}^{n} c_{i}(\boldsymbol{\lambda}) e_{i}(t) \right\Vert^2 \, ,
\end{equation}

where $\left\Vert \cdot \right\Vert$ is the $L_2$ norm. To find the reduced basis
representation we use a greedy algorithm implemented in the {\tt rompy} python
package~\cite{rompy,PhysRevX.4.031006}. We begin by densly sampling the
parameter space and thus creating our training set $\mathcal{T}_{TS}$, we then
pick one of the points randomly to seed the greedy algorithm. This seed point is
the first \emph{greedy point} and the first element in the basis $B$. The greedy
algorithm iteratively builds up the basis by computing the current
representation error $\sigma$ against all points in $\mathcal{T}_{TS}$. The
sample with the largest representation error is added to the set of greedy
points and also added to the basis using the iterative-modified Gram-Schmidt
algorithm~\cite{Hoffmann2005IterativeAF}. The greedy algorithm stops when the
sample with the largest representation error is already in the basis or if the
largest representation error is below the user specified tolerance
$\sigma_{\rm{tol}}$. This results in a set of $m$ greedy points and a basis $B$
of size $m$ that covers $\mathcal{T}_{TS}$ to within an accuracy of
$\sigma_{\rm{tol}}$. If the $\mathcal{T}_{TS}$ is sufficiently dense and thus
representative of the entire $\mathcal{T}$ then we can use the reduced basis to
approximate the function for any point in $\mathcal{T}$.

After we have built a reduced basis we use the empirical interpolation method
(EIM)~\cite{BARRAULT2004667, 1534-0392_2009_1_383} to construct an empirical
interpolant of $X(t; \boldsymbol{\lambda})$. This results is a new basis,
$\bar{B}_n = \{e_i(t)\}_{i=1}^{n}$, also of size $m$ that is constructed such
that the coefficients of the basis
$\{\alpha_j(\boldsymbol{\lambda})\}_{j=1}^{n}$ are values of the function $X$
themselves at the empirical time nodes $T_j$

\begin{equation}
\alpha_j(\boldsymbol{\lambda}) = X(T_j; \boldsymbol{\lambda}) \, .
\end{equation}

In order to evaluate the surrogate model at any point in $\mathcal{T}$ you can
approximate the $\alpha$ coefficients by either fitting or interpolating them
across $\mathcal{T}$. We denote the fitted coefficients as $\hat\alpha$. We use
the EIM because typically the variation of the $\alpha$ coefficients is smoother
than the $c$ reduced basis coefficients. This makes it easier to fit or
interpolate the coefficients and requires a smaller training set to obtain a
model of the coefficients.

Up to this point we only know the $\alpha$ coefficients at the greedy points.
This is typically not enough points to sample the $\alpha$ functions to
accurately fit or interpolate across the parameter space. In this paper we build
a surrogate model of SEOBNRv4, which permits us to generate large training
sets that we can use to sample the $\alpha_j(\boldsymbol{\lambda})$ functions.

Finally, the surrogate model for $X(t; \boldsymbol{\lambda})$ is defined as

\begin{equation}
X^S(t; \boldsymbol{\lambda}) \approx \sum_{i=1}^{n} \hat{\alpha}_{i}(\boldsymbol{\lambda}) \bar{B}_{i}(t) \, .
\label{equ:surfunc}
\end{equation}

The problem therefore, is reduced to finding a fast and accuracte approximation
to $\alpha$. There are many machine learning methods that can be used to
interpolate or fit these coefficients and depending on the accuracy required and
the dimensionality different methods will be more suitable than others.
In~\cite{2014CQGra..31s5010P} the authors interpolate the reduced basis coefficients
directly in the 3D aligned-spin parameter space. Interpolation is a good method
for low dimensional parameter spaces but in dimensions $\gtrsim 3$ interpolation
becomes difficult due to the large number of data points typically required.
In~\cite{PhysRevD.95.104023,PhysRevD.99.064045} the authors built a surrogate
model for numerical relativity produced precessing BBHs corresponding to a 7D
parameter space with the EIM. Here, due to the relatively small size of their
training set and the high dimensional parameter space interpolation was not
appropriate and instead used a basis of monomials constructed with a greedy
algorithm to reduce overfitting. In~\cite{PhysRevD.99.064045} the authors
modelled the 3D aligned-spin parameter space with numerical relativity
simulations using EIM and fit the coefficients using Gaussian process
regression.

In~\cite{Setyawati:2019xzw} the authors systematically explored several methods and
ranked them in terms of accuracy, time to fit and prediction time. The also
experimented with ANNs but restricted to shallow networks with only 2 hidden
layers and training/execution on CPUs only.
In~\cite{PhysRevLett.122.211101} the authors modelled the
reduced basis coefficients of post-Newtonian inspiral waveforms using a 4D
parameter space, comprised of the component masses and the aligned-spin
components, using ANNs. In this work we use a similar approach but instead
applied to the empirical interpolation (EI) $\alpha$ coefficients and model the
complete inspiral, merger and ringdown signal.

\section{binary black hole surrogate model}
\label{sec:bbhmodel}

\subsection{parameter space}

In this paper we investigate the possibility to use ANNs in the construction of
surrogate waveform models for BBH signals. We build a surrogate model of the
model SEOBNRv4~\cite{2017PhRvD..95d4028B}. It predicts the GW signal emitted
from non-eccentric BBH mergers where the black hole spin angular momenta are
constrained to be parallel (or antiparallel) with the orbital angular momentum.
Extensions of this model to include subdominant multipoles and precession have
been done~\cite{2018PhRvD..98h4028C, 2020arXiv200409442O} however, we develop
our method with the simpler case. This model is based on the effective-one-body
(EOB) formalism, extened to predict the merger and ringdown signal by fitting
free coefficients to numerical relativity solutions. This is a time domain model
where the inspiral model is calculated by solving the EOB Hamiltonian equations
of motion; a set of coupled, ordinary differential equations. This method has
proven to provide accurate GW templates but typical implementations of EOB
models tend to be computationally expensive. As such a lot of work has gone into
optimising the production of EOB templates; either by improving the
computational efficiency of the inspiral
calculation~\cite{Knowles:2018hqq,Devine2016OptimizingST,PhysRevD.99.021501} or
by developing, frequency domain, reduced order surrogate
model~\cite{2014CQGra..31s5010P}. Whilst there already exists frequency domain
surrogate models for both SEOBNRv4~\cite{2017PhRvD..95d4028B} and
SEOBNRv4HM~\cite{2020PhRvD.101l4040C} it is an excellent model to develop new
methodology and we apply this to the time domain rather than the frequency
domain.

Motivated by past work~\cite{PhysRevD.99.064045} and to facilitate comparisons
we build a surrogate model of SEOBNRv4 covering mass-ratios from 1:1 to
1:8 and allowing each BH spin to range from $-0.99$ to $0.99$. For each systems
we generate the $h_{22}(t) \in \mathbb{C}$ multipole data. This method to
construct the reduced basis requires that all data are evaluated on the same
time grid. We choose to build a surrogate model that is valid from $15
\,\rm{Hz}$ at a total mass of $60 M_\odot$ for all mass-ratios and spins in the
training set. To find the start frequency of the surrogate, $f_{\rm{start}}$,
for a new total mass, $M_{\rm{new}}$, we can use the formula
$f_{\rm{start}}(M_{\rm{new}}) = 15 \times (60 M_\odot/M_{\rm{new}}) \, \rm{Hz}$.
We work with geometric units $M$ and perform a time shift such that $t=0 M$
corresponds to the peak of the amplitude.
It is important that this procedure is done with high accuracy to avoid
an unnecessarily large reduced basis~\cite{PhysRevX.4.031006}.

To ensure that the surrogate is valid for the domain stated above we generate
all waveforms with a lower start frequency of $8 \,\rm{Hz}$ and then truncate
all data such that the data starts at at least $15 \,\rm{Hz}$. For the parameter
space we consider this corresponds to a start time of $-20000 M$. In addition
to performing a time shift to the data we also perform a phase shift such that
the phase is zero at the start time (i.e., $-20000 M$). We keep $100 M$ of
post-peak ringdown data. Finally the data is resampled at a resoltuion of
$\Delta t = 0.5 M$ onto the domain $[t_{\rm{min}}, t_{\rm{max}}] = [-20000, 100] M$.
When analysing long signals then the physical constraints of computer memory
becomes an issue. There are a number of ways to compress the training data which
typically involve non-uniformly sampled data~\cite{PhysRevD.94.044031,
2016arXiv161107529G, 2017CQGra..34k5006V, 2020arXiv200110897G} however,
these were unnecessary here.

We generate three different sets of data: training, validation and test sets.
The training set is used to 1) build the reduced basis and 2) densely sample the
projection coefficients that we will fit. The validation set is also used to
sample the projection coefficients but is only used to monitor the accuracy of
the fit to diagnose if the model is under- or over- fitting and to help tune the
hyperparameters of the network. The test set, or hold-out set, is not used in
the training of the network but is used to evaluate the final accuracy of the
model. The validation and test sets serves as a way to assess the size
``generalisation gap'' of the model, the distance between the performance of the
model on the training set and on the hold-out set. The training set contains $2
\times 10^5$ samples and the validation and test set both contain $2 \times 10^4$
samples.

\subsection{waveform performance metrics}

To quantify the level of agreement between two, real-valued, waveforms $h_1$ and
$h_2$ we use the standard inner product weighted by the noise power spectral
density (PSD) of the GW detector $S_n(f)$. It is defined
as~\cite{Cutler1994GravitationalWF}

\begin{equation}
\langle h_1, h_2 \rangle = 4 \, {\rm{Re}} \int_{f_{\rm{min}}}^{f_{\rm{max}}} \frac{\tilde{h}_{1}(f) \tilde{h}_{2}^{*}(f)}{S_n(f)} df \, .
\end{equation}

The match between two waveforms is defined as the inner product between
normalised waveforms $(\hat{h} \equiv h / \sqrt{\langle h,h \rangle})$
maximised over a relative time $(t_0)$ and phase $(\phi_0)$ shift between the two waveforms,

\begin{equation}
M(h_1, h_2) = \max_{t_0, \, \phi_0} \langle \hat{h}_1, \hat{h}_2 \rangle \, .
\end{equation}

Finally we shall quote results in terms of the mismatch which is
the fractional loss in the signal-to-noise ratio due to
modelling errors defined as

\begin{equation}
    \mathcal{M}(h_1, h_2) = 1 - M(h_1, h_2) \, .
\label{equ:mismatch}
\end{equation}

\subsection{reduced basis construction}

We choose to monitor the relative greedy error, that is the error relative to
the representation error at the first iteration. To determine what value to use
for the tolerance we varied the tolerence from $10^{-6}$ to $10^{-16}$
logarithmically in steps of $2$. For each resulting basis we computed the
mismatch (equation~\ref{equ:mismatch}) between the training data and the basis
representation. We also did this for the validation set and Table~\ref{tab:tol}
shows the results. We find that the number of basis functions grows much faster
for the amplitude than for the phase.

We base our choice of greedy tolerance, and therefore on the number of basis
functions to use, on the accuracy of the SEOBNRv4 model.
In~\cite{2017PhRvD..95d4028B} the accuracy in terms of the mismatch was found to
be between $10^{-2}-10^{-4}$ when compared to numerical relativity data.
Therefore, we use a greedy tolerance of $10^{-10}$, which produces a basis with
mismatch errors of at worst $\sim 6.5 \times 10^{-5}$ for both the training and
validation set. The consistency between the training and validation set implies
that we have sampled the space with the training set densly enough that the
basis can represent out of sample waveforms with equivalent accuracy. This
produces a reduced basis with only 19 basis functions for the amplitude and 8
basis functions for the phase.

\begin{table}[]
\begin{tabular}{lllll}
\hline
\multicolumn{1}{c}{\begin{tabular}[c]{@{}c@{}}Greedy\\ Tolerance $\sigma_{\rm{tol}}$\end{tabular}} & \multicolumn{1}{c}{Training Set} & \multicolumn{1}{c}{Validation Set} & \multicolumn{1}{c}{\begin{tabular}[c]{@{}c@{}}\# Bases:\\ Amplitude\end{tabular}} & \multicolumn{1}{c}{\begin{tabular}[c]{@{}c@{}}\# Bases:\\ Phase\end{tabular}} \\ \hline
$10^{-6}$                                                                      & $3.1 \times 10^{-1}$              & $3.1 \times 10^{-1}$                & 9                                                                                 & 3                                                                             \\
$10^{-8}$                                                                      & $1.9 \times 10^{-2}$              & $1.8 \times 10^{-2}$                & 13                                                                                & 5                                                                             \\
$10^{-10}$                                                                     & $6.5 \times 10^{-5}$              & $6.2 \times 10^{-5}$                & 19                                                                                & 8                                                                             \\
$10^{-12}$                                                                     & $1.2 \times 10^{-6}$              & $1.1 \times 10^{-6}$                & 39                                                                                & 12                                                                            \\
$10^{-14}$                                                                     & $1.1 \times 10^{-8}$              & $8.2 \times 10^{-9}$                & 91                                                                                & 33                                                                            \\
$10^{-16}$                                                                     & $9.7 \times 10^{-10}$             & $9.7 \times 10^{-10}$               & 102                                                                               & 51                                                                            \\ \hline
\end{tabular}
\caption{Worst mismatch of the reduced basis and reduced basis size (for amplitude and phase bases) as a function of greedy error tolerance.}
\label{tab:tol}
\end{table}

\section{Neural Network Training Strategy}
\label{sec:nnts}

In this section we investigate how different choices of data pre-processing,
neural network architecture, optimizers and mini-batch size impact the networks
ability to fit (or learn) the data. We will call the combined set of choices our
\emph{training strategy} and our goal is to find the optimal training strategy
to minimise the loss function over different training strategies. We only
outline our investigation here and leave details to
appendix~\ref{app:nn-explore}.

In general is it not trivial to know how a particular change to any of these
parameters will effect the network or indeed if the choices are independent of
each other. In order to make this problem tractable we will use a greedy method,
making localy optimal choices at each step. We explore each aspect of the
training strategy in the following order: (i) data pre-processing, (ii) width
and depth of the neural network, (iii) activation functions and finally (iv)
optimizers. At each step we perform the experiment twice, once using batched
gradient decent (using the entire dataset) and again using mini-batch gradient
decent with a mini-batch size of 1000~\cite{10.5555/3086952}. At each step we
typically will take the neural network which has the smallest final loss as use
those parameters in the next step however, in some tests we find there are
several network configurations that perform equally well. For those cases we
used the settings that resulted in the fastest trained network. We note that if
the ordering of exploration was different then it is possible that we would end
up with a different training strategy.

The independent variables of the data we will fit are; the mass-ratio ($q$), the
aligned-spin component of the primary ($\chi_{1}$) and the aligned-spin
component of the secondary ($\chi_{2}$). As done in previous surrogate
models~\cite{PhysRevD.99.064045,2020PhRvD.101h1502R} we first perform a
logarithmic transformation on the mass-ratio, as we also find that this helps
fit the data more accurately. In the following sections we will refer to the
independent variables i.e., $\log(q)$, $\chi_{1}$ and $\chi_{2}$ simply as \X
and the dependent variables i.e., the coefficients of the empirical
interpolation basis as \Y. For the amplitude \Y is a 19 dimensional vector and
for the phase it is a 8 dimensional vector (see Table~\ref{tab:tol}).

We use {\tt TensorFlow}~\cite{tensorflow2015-whitepaper} and {\tt
Keras}~\cite{chollet2015keras} to design and train two independent feed-forward,
fully-connected neural networks, one for the amplitude and one for the phase,
using the mean-squared error loss function. The input layer is given by the
dimensionality of the independent variables (\X). The rest of the network;
number of hidden layers, number of neurons in each layer and choice of
activation function will be explored. The output layer uses a linear activation
function (suitable for regression problems) and the number of output neurons is
given by dimensionality of the dependent variables (\Y). We train the networks
using the backpropagation algorithm to minimise the loss function with respect
to the network's weights and biases.

One of the key decicions to make is how should you choose the learning rate
for the stochastic gradient decent algorithm.
Some authors suggest that the choice of mini-batch size should be linked with
the choice of learning rate~\cite{2017arXiv170508741H, Granziol2020CurvatureIK}.
We explore a range of different optimizers in appendix~\ref{app:nn-explore}
but always use a learning rate that decreases with time according to

\begin{equation}
\tau_{k} = (\tau_{\rm{init}} - \tau_{\rm{final}}) / (1 + R \floor{k/\Delta k}) + \tau_{\rm{final}} \, .
\label{eq:lrs}
\end{equation}

Where $\tau_{k}$ is the learning rate at epoch $k$, $\tau_{\rm{init}}$ is the
initial learning rate ($10^{-3}$), $\tau_{\rm{final}}$ is the final learning
rate ($10^{-5}$), $R$ is the decay rate ($10$) and $\Delta k$ is the interval
between decaying ($2000$), unless otherwise stated the values we use are given
in parentheses. We choose to compute the floor of the ratio $k/\Delta k$ which
means the learning rate exhibits steps-wise changes. Some optimisers, such as
Adam~\cite{Kingma2015AdamAM}, already use an adaptive learning rate, however by
using a learning rate scheduler we can futher control the maximum value of the
learning rate as a function of time (epoch).

\subsection{Final Neural Network Model}

The final training strategies for the amplitude and phase data are given in
Table~\ref{tab:nn}. The networks were for trained for $10^5$ epochs with a
mini-batch size of $1000$ which took $\sim 6-7$ hours on a Tesla P100 GPU.

We find that the data pre-processing method had a large impact on the
performance of the networks, see appendix~\ref{app:nn-explore-preproc} for
details. For the amplitude data, the optimal pre-processing methods are to
normalize the \X data and use the raw \Y data. For the phase we normalize the \X
data and scale the \Y data. For both the amplitude and phase networks we use 4
hidden layers, each with a width of 320 units per layer. As detailed in
appendix~\ref{app:nn-explore} we find that deeper networks can achieve lower
losses but not by a significant amount. For the hidden layer activation
functions we find that the ReLU function performed best for the amplitude data
and the Softplus function performed best for the phase data. Finally we used the
Adam optimizer for the amplitude data and the AdaMax optimizer for the phase
data.

\begin{figure}[t]
\includegraphics[width=\linewidth]{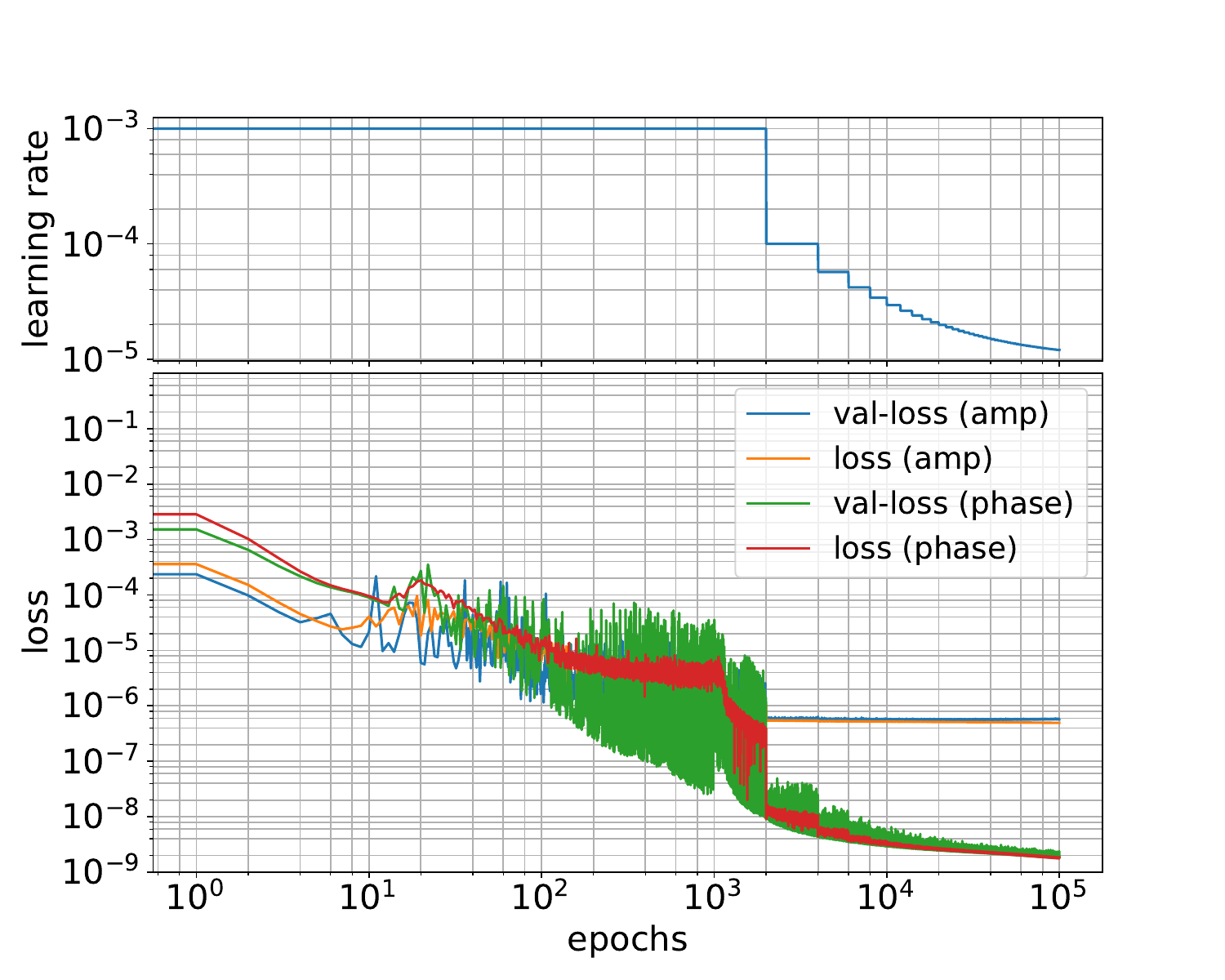}
\caption{Top panel: learning rate. Bottom panel: The amplitude [phase] loss
(orange [red]) and validation loss (blue [green]) curves as a function of
epochs.}
\label{fig:final-nn-loss}
\end{figure}

In Figure~\ref{fig:final-nn-loss} (bottom panel) we show the loss and
validation-loss learning curves for the amplitude and phase data on a log-log
scale. The top panel shows the learning rate as a function of epoch, which
decreases according to Equation~\ref{eq:lrs}, every 2000 epochs. The sudden
drops in the loss curves correspond to the drops in the learning rate.

We find that the amplitude data shows some very mild signs of over-fitting and
the phase data shows signs of under-fitting however, as we will see in the next
section, these networks produce mismatch errors below our error tolerance.

\begin{table}
\begin{tabular}{lll}
\hline
~                   & Amplitude       & Phase           \\ \hline
\X pre-processing   & Standard Scaler & Standard Scaler \\
\Y pre-processing   & None            & Min-Max Scaler  \\
N-hidden Layers     & 4               & 4               \\
Units per layer     & 320             & 320             \\
Activation Function & ReLU            & Softplus        \\
Optimizer           & Adam            & AdaMax          \\
Final Loss          & 4.93e-07        & 1.82e-09        \\
Final Val-Loss      & 5.74e-07        & 1.90e-09        \\
Training Time       & $6-7$ hrs       & $6-7$ hrs       \\
\hline
\end{tabular}
\caption{Final training strategy for amplitude and phase data. Data
pre-processing, Neural Network architecture and hyper parameter choices for
amplitude and phase data. A mini-batch size of 1000 was used for both.
GPU used: Tesla P100.}
\label{tab:nn}
\end{table}

\section{model evaluation}

With the final neural network models for the EI amplitude and phase coefficients
in hand we can evaluate the performance of the neural network powered surrogate
model (ANN-Sur) we have built to mimic SEOBNRv4. We scrutinize the surrogate
model using a two different tests. The first test
(section~\ref{subsec:mismatch}) is to see how accurate the surrogate model is
when compared to the original model. The second test
(section~\ref{subsec:speed}) is to quantify what is the speed improvement we
have achieved compared with SEOBNRv4. We also compare to other state-of-the-art
models in terms of computational efficiency and the improvement obtained when
running the model on a GPU rather than a CPU.

\subsection{Mismatch vs total mass}
\label{subsec:mismatch}

\begin{figure}[t]
\includegraphics[width=\linewidth]{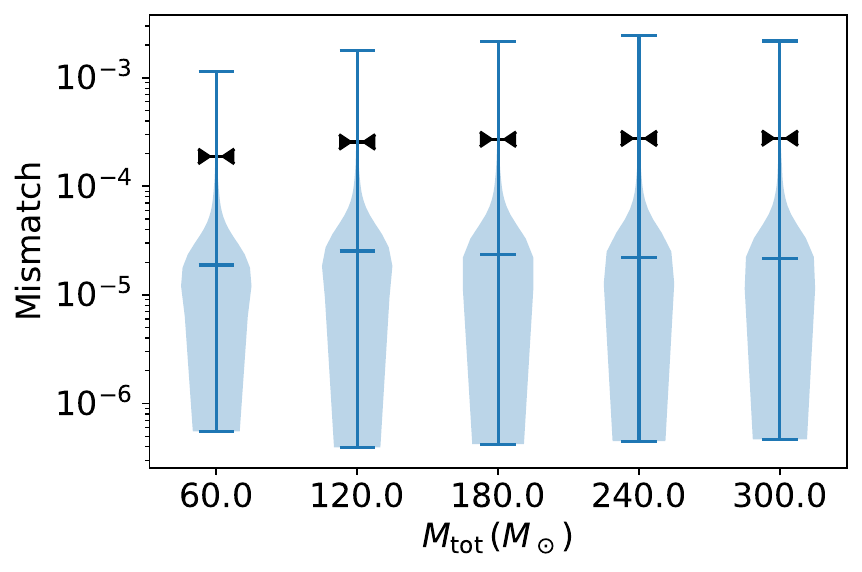}
\caption{Mismatches between the ANN-Sur and SEOBNRv4 validation dataset
represented as a violin plot. The median is marked by the middle horizontal
line and the extent of the lines show the minimum and maximum values. The
envelope is proportional to the density of points. The black triangles mark the
95th percentile. We remind the reader that the accuracy of the SEOBNRv4 model is
between $10^{-2}-10^{-4}$~\cite{2017PhRvD..95d4028B}.}
\label{fig:mismatches-mtot}
\end{figure}

\begin{figure}[t]
\includegraphics[width=\linewidth]{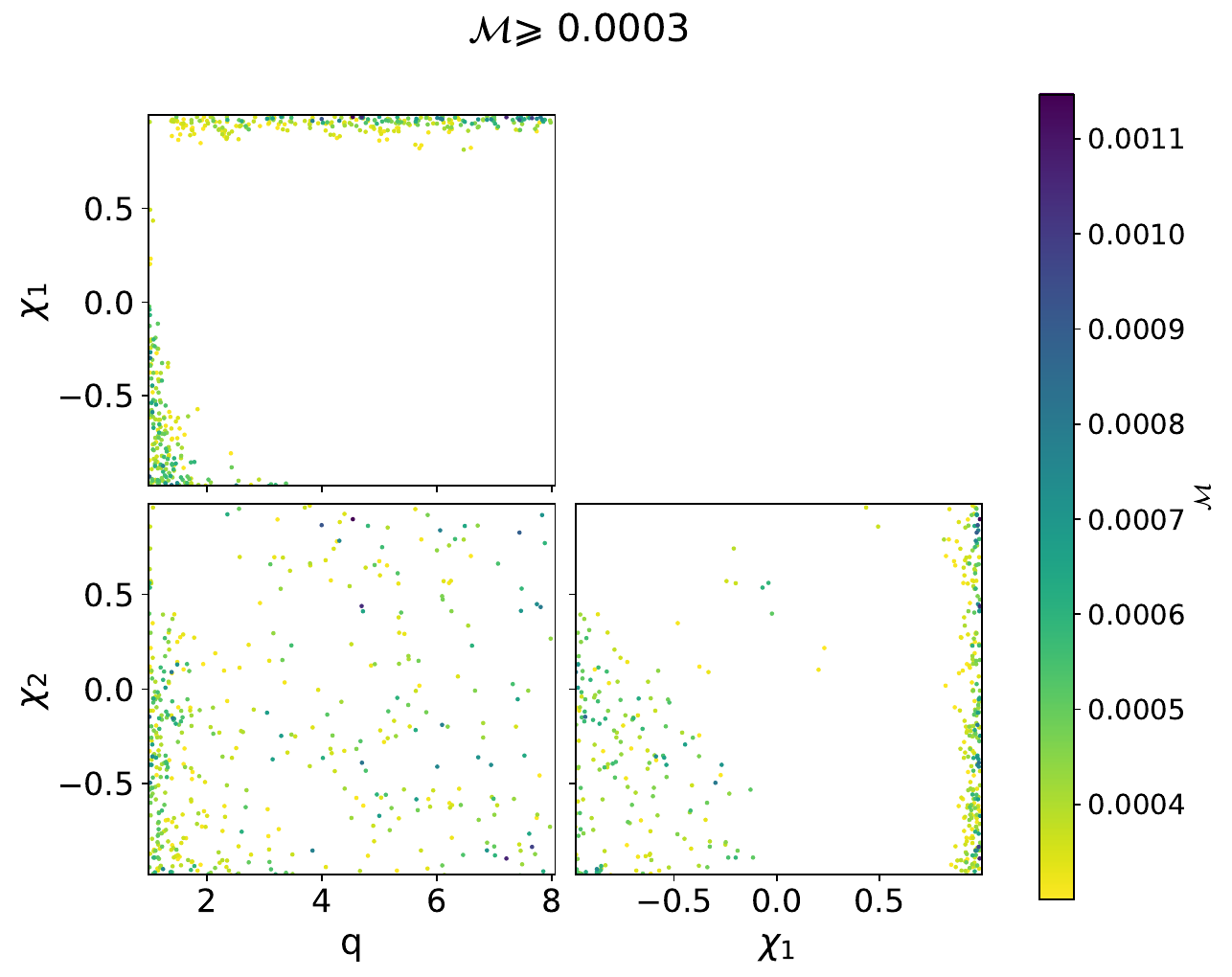}
\caption{Mismatches plotted across the $(q, \chi_1, \chi_2)$ parameter space.
Only cases with mismatches larger than the 95th percentile ($\expnumber{3}{-4}$)
are shown. This is the result for $M_{\rm{tot}} = 60 M_\odot$ but other
$M_{\rm{tot}}$ are similar.}
\label{fig:corner-mismatches}
\end{figure}

To quantify the accuracy of the surrogate model we compute the mismatch, using
the expected noise curve for Advanced LIGO operating at design
sensitivity~\cite{aligozerodethpnew}, between ANN-Sur and all the waveforms in
the validation dataset noting that results are similar for the training and test
datasets. Due to the shape of the PSD the smaller (larger) values of
$M_{\rm{tot}}$ tend to accentuate modelling errors during the inspiral (merger)
therefore, we consider the following values for $M_{\rm{tot}} = ( 60, 120, 180,
240, 300 ) M_\odot$. We used a low frequency cut-off of $15\,{\rm{Hz}}$ and
variable high frequency cut-off given by $1.4 f_{\rm{RD}}\,{\rm{Hz}}$ where
$f_{\rm{RD}}$ is an estimate of the final BH ringdown
frequency~\cite{Husa2016FrequencydomainGW}. The results of which are shown in
Figure~\ref{fig:mismatches-mtot}. We find that the mismatch is stable as a
function of $M_{\rm{tot}}$ with a slight rise in the mismatch by
$\expnumber{1}{-3}$ for larger values of $M_{\rm{tot}}$. The vast majority of
cases have mismatches below $\sim \expnumber{3}{-4}$ (95th percentile) with a
median value of $\sim \expnumber{2}{-5}$. The lowest mismatch we achieve is
$\sim \expnumber{4}{-6}$. The highest mismatch obtained is $\sim
\expnumber{2}{-3}$ and these cases are distributed primarily in two clusters as
shown in Figure~\ref{fig:corner-mismatches}. One cluster is towards the upper
boundary of $\chi_1$. The other cluster is towards corner of low $\chi_1$ and
low $q$. If more training points in these regions do not improve performance
here then a domain decomposition strategy can be employed.

\subsection{Computational Speed}
\label{subsec:speed}

Most waveform models are designed to run on CPUs with some recent work on moving
waveform generation onto a
GPU~\cite{Chua2020LearningBP,Katz2020GPUacceleratedMB,Talbot2019ParallelizedIF,2020arXiv200806071C}
however, it is still an open question of how waveform generation can make the
most use of GPUs. With {\tt TensorFlow} we can generate optimized {\tt
TensorFlow} graphs with accelerated linear algebra (XLA)~\cite{xla} compilation
that can be executed on either a CPU or GPU. Here we used an Intel 2.20GHz Xeon
CPU E5-2630v4 and a TITAN X Pascal GPU for our comparisons.

In Table~\ref{tab:speed-comparison} we quantify the speed-up we achieve compared
with the original SEOBNRv4 model as well as the optimized version of the model
SEOBNRv4opt. We generated the GW signal with the following parameters $q=3$,
$M_{\rm{tot}}=60 M_\odot$, $\chi_{1} = 0.8$, $\chi_{2} = 0.5$, use a sample rate
of $1/2048 \, {\rm{s}}$ and an initial frequency of $f_{\rm{min}} = 12 \,
{\rm{Hz}}$ (corresponds to a length of $\sim 20000 \, M$). We find that the
SEOBNRv4 model takes $1812 \, {\rm{ms}}$ to compute this waveform with the
SEOBNRv4opt model improving upon this by a factor of $\sim 20$ to $91.6 \,
{\rm{ms}}$. The ANN-Sur model on a CPU takes $2.7 \, {\rm{ms}}$ giving a speed
up of $661$ ($33$ with respect to SEOBNRv4opt). When running the ANN-Sur model
on a GPU waveform generation takes just $0.4 \, {\rm{ms}}$ giving a speed up of
$4646$ ($235$ with respect to SEOBNRv4opt).

We generated the same GW signal with other state-of-the-art GW signal models for
the dominant (2,2) harmonic for non-precessing binaries. SEOBNRv4ROM is also a
surrogate model for SEOBNRv4, however it is constructed in the frequency domain
and interpolates reduced basis projection coefficients.
NRHybSur3dq8~\cite{PhysRevD.99.064045} is a time domain surrogate model for
numerical relativity simulations produced with the SpEC code and hybridised with
PN/EOB inspiral waveforms. It also uses EIM but models the $\alpha$ projection
coefficients using Gaussian Process Regression.
IMRPhenomD~\cite{PhysRevD.93.044007} and it's successor
IMRPhenomXAS~\cite{Pratten2020SettingTC} are frequency domain phenomenological
models. Phenomenological models combine results from post-Newtonian theory,
black hole perturbation theory and numerical relativity solutions together with
sophistocated modelling techniques to build bespoke models for the GW signal. We
note that comparing to other surrogate models should be done with caution. The
computationally speed-up of a surrogate model comes from (i) the size of the
basis and (ii) the efficiency of the method used to estimate the projection
coefficients. Both of these are effected by the parameter space (including the
duration of the signal) that the surrogate hopes to cover. Therefore, for
SEOBNRv4ROM and NRHybSur3dq8, that cover longer duration signals, the
comparisons relate to their specific implementation and not necessarily to the
optimal performance of the method used to predict the basis coefficients.

We find that NRHybSur3dq8 takes the longest to generate this waveform taking $38.6
\, {\rm{ms}}$. Next, SEOBNRv4ROM taking $6.9 \, {\rm{ms}}$. Finally, the
fastest models are the IMRPhenomD and IMRPhenomXAS models taking $\sim 1.2 \,
{\rm{ms}}$. ANN-Sur is highly competetive in terms of
computational speed, outperforming all but the IMRPhenom models when run on a
CPU and outperforms all models when run on a GPU by a factor of $\sim 3$.

Some calculations can be rapidly accelerated by using a GPU by processing
similar calculations in parallel using \emph{batches}. ANN-Sur is built with
{\tt TensorFlow} and can readily take advantage of this. In
Table~\ref{tab:speed-cpu-gpu} we time how long ANN-Sur takes to generate random
batches of (10, 100, 1000, 10000) waveforms, averaged over 100 trials, both on a
CPU and a GPU. We find that even on a single CPU the batched calculation can
produce $\expnumber{1}{4}$ waveforms in $\sim 5 \, {\rm{s}}$ and the use of a
GPU provides a speed-up factor of $\sim 30$ taking only $164.3 \, {\rm{ms}}$. To
generate the same number of SEOBNRv4opt waveforms on a single CPU we estimate it
would take $\sim 15 {\rm{mins}}$. Therefore, ANN-Sur produces a speed-up factor
of $\sim 5550$.

The ability to extremely efficiently produce large numbers of template waveforms
simultaneously on a single CPU or GPU has the potential to substantially reduced
the computational cost of GW analyses such as parameter
estimation~\cite{Pankow2015NovelSF,Lange2018RapidAA} and in the generatation of
GW template banks~\cite{Harry2016SearchingFG,Roy2019EffectualGT}.

\begin{table}
\begin{tabular}{lll}
\hline
Model & Time (ms) & Speed-up \\ \hline
SEOBNRv4 (opt) & 1812 (91.6)  & -  \\
ANN-Sur CPU & 2.7 & 661 (33) \\
ANN-Sur GPU & 0.4 & 4646 (235) \\ \hline
*SEOBNRv4ROM & 6.9 & - \\
*IMRPhenomD & 1.2 & - \\
*IMRPhenomXAS & 1.3 & - \\
NRHybSur3dq8 & 38.6 & - \\
\hline
\end{tabular}
\caption{Average time (ms) to generate a one waveform averaged over 100
waveforms. Times and speed-ups in parentheses correspond to the SEOBNRv4opt
model. $q=3$, $M_{\rm{tot}}=60 M_\odot$, $\chi_{1} = 0.8$, $\chi_{2} = 0.5$.
$f_{\rm{min}} = 12 \, {\rm{Hz}}$ (corresponds to a length of $\sim 20000 \, M$).
For time-domain approximants we used a sample rate of $1/2048 \, {\rm{s}}$.
Models prefixed with a * are frequency-domain models and we used a sample rate
of $1/8 \, {\rm{Hz}}$. When evaluating ANN-Sur, SEOBNRv4ROM and NRHybSur3dq8 ``warm up'' execution is performed to load one-time overhead data.
Additionally for NRHybSur3dq8 we only evaluate the (2, 2) mode.}
\label{tab:speed-comparison}
\end{table}

\begin{table*}
\centering
\begin{tabular}{llllll}
\hline
& \multicolumn{2}{c}{CPU}                                                           & \multicolumn{2}{c}{GPU}                                                           & \multirow{2}{*}{\begin{tabular}[c]{@{}l@{}} \, Speed-up\\(CPU/GPU)\end{tabular}}  \\
\cline{2-5}
& Total Time (ms) & \begin{tabular}[c]{@{}l@{}}Time Per\\Waveform (ms)\end{tabular} & Total Time (ms) & \begin{tabular}[c]{@{}l@{}}Time Per\\Waveform (ms)\end{tabular} &                             \\
\hline
Single        & 2.7             & 2.7                                                             & 0.4             & 0.4                                                             & 7                           \\
Batched ($10$) & 13              & 1.3                                                             & 0.5             & 0.05                                                            & 26                          \\
Batched ($10^2$) & 73.3            & 0.73                                                            & 2.1             & 0.021                                                           & 35                          \\
Batched ($10^3$) & 575.4           & 0.58                                                            & 16.98           & 0.017                                                           & 34                          \\
Batched ($10^4$) & 5010            & 0.50                                                            & 163.4           & 0.016                                                           & 31                          \\
\hline
\end{tabular}
\caption{Computational efficiency of ANN-Sur when generating batches of
waveforms.}
\label{tab:speed-cpu-gpu}
\end{table*}

\section{Conclusion}
\label{sec:conclusion}

In the next five years the size of GW catalogues is expected to grow from
$\mathcal{O}(10)$ to $\mathcal{O}(10^3)$~\cite{Baibhav2019GravitationalwaveDR,
Aasi:2013wya}. It is therefore imperitive that we device methods that can make use of
the most accurate waveform models, which are typically also the most
computationally expensive, in the analysis of all GW events.

In this paper we have presented ANN-Sur, our methodology to construct surrogates
for GW signal models powered by artificial neural networks. A similar idea was
presented in~\cite{PhysRevLett.122.211101} with a focus on inspiral-only signal
models and masses suitable for LISA detector. Here we focus on GW signals for
the complete inspiral, merger and ringdown with a mass range targeted for
current ground-based detectors. As a first application of our method we have
built a time-domain surrogate model of the SEOBNRv4 model for spin-aligned
binary black hole mergers, which covers the following 3D intrinsic parameter
space: $q \in [1, 8]$, $\chi_{1,2} \in [-0.99, 0.99]$. We built the surrogate to
be valid from $15$ Hz for a total mass of $60 M_\odot$, which leads to a length
of $\sim 20000 \, M$. When compared with the original SEOBNRv4 model our
surrogate model has a worst mismatch of $\sim \expnumber{2}{-3}$ and a median
mismatch of $\sim \expnumber{2}{-5}$, see Figure~\ref{fig:mismatches-mtot}.

ANN-Sur is built with the {\tt TensorFlow} library and can seamlessly run on
either a CPU or GPU. In section~\ref{subsec:speed} we compared the computational
efficiency of ANN-Sur with the original SEOBNRv4 model. We find that the average
time to compute a single waveform with the optimised SEOBNRv4 model is $91.6 \,
{\rm{ms}}$, when running ANN-Sur on a CPU this is reduced to $2.7 \, {\rm{ms}}$
and when run on a GPU takes just $0.4 \, {\rm{ms}}$, a factor of 235
improvement. When comparing with the frequency-domain surrogate model
SEOBNRv4ROM we find that ANN-Sur is a factor of 2.5 (17) times faster when run
on a CPU (GPU). We expect that frequency-domain surrogate models built using
this method would be significantly improved, which may further increase the
performance of likelihood acceleration techniques such as the reduced order
quadrature
rule~\cite{Canizares2015AcceleratedGW,PhysRevD.94.044031,Morisaki2020RapidPE}.

ANN-Sur also permits us to generate large numbers of waveforms simultaneously in
batches on a single CPU or GPU. In Table~\ref{tab:speed-cpu-gpu} we find that we
can generate batches of up to $10^{4}$ waveforms in $\sim 5 \, {\rm{s}}$ on a
CPU and in just $\sim 160 \, {\rm{ms}}$ on a GPU, corresponding to a per
waveform generation time of just $0.016 \, {\rm{ms}}$. This new kind of
parallelisation allows for the generation of large training sets to train deep
learning methods to perform Bayesian
inference~\cite{Chua2020LearningBP,Gabbard2019BayesianPE,Green2020CompletePI} or
to rapidly generate waveforms for grid-based methods suchs as
~\cite{Pankow2015NovelSF,Lange2018RapidAA,Wysocki2019AcceleratingPI}. The
increased computational efficiency gained here should also be obtained for
binary neutron star systems~\cite{Dietrich2019MatterII,Nagar2018TimedomainEG}
and neutron star black hole
binaries~\cite{Thompson2020ModelingTG,Matas2020AnAN}, increasing the likelihood
that we will find multimessenger events~\cite{GBM:2017lvd}.

Whilst our surrogate meets current accuracy requirements, with only 19 and 8
basis functions for the amplitude and phase respectively, higher accuracy
surrogate models will be required in the future as detectors become more
sensitive. Higher accuracy surrogates can be built by including more basis
functions, for example see Table~\ref{tab:tol}, however, we found that the ANNs
we used were unable to model the projection coefficients accurately enough. This
issue should be solved by using larger training sets and improving our training
strategy.

One of the next steps will be to incorporate the full BBH parameter space i.e.,
build a surrogate model that includes spin-precession and higher
harmonics~\cite{PhysRevResearch.1.033015,PhysRevD.96.024058,2020arXiv200409442O,Khan2020IncludingHO,Pratten2020LetsTA}.
Extending our method to work effectively in higher dimensions is also possible
by increasing the size of the training set and network capacity.

A final and unique advantage of our method is to be able compute waveform
derivatives using automatic differentiation~\cite{2018arXiv181105031M}. This is
a key ingredient for the Bayesian inference sampling method Hamiltonian Monte
Carlo (HMC)~\cite{DUANE1987216,Neal2011MCMCUH,Betancourt2017ACI}. This has
rarely been used in the GW astronomy community
~\cite{Porter2014AHM,Bouffanais2018BayesianIF} as the computational cost of
computing the required likelihood derivatives quickly offsets any performanced
gained from using HMC. We are currently exploring the benefits of combining HMC
with ANN-Sur which will be presented in the future~\cite{GreenKhanInPrep}.

\begin{acknowledgments}

We thank Alvin Chua, Edward Fauchon-Jones, Vasileios Skliris, Michael Norman,
Luke Berry, David Sullivan, Vivien Raymond and Mark Hannam for useful
discussions.
S.K. was supported by European Research Council Consolidator Grant 647839.
R.G. was supported by Science and Technology Facilities Council (STFC) grant
ST/L000962/1.
This work was performed using the Cambridge Service for Data Driven Discovery
(CSD3), part of which is operated by the University of Cambridge Research
Computing on behalf of the STFC DiRAC HPC Facility (www.dirac.ac.uk). The DiRAC
component of CSD3 was funded by BEIS capital funding via STFC capital grants
ST/P002307/1 and ST/R002452/1 and STFC operations grant ST/R00689X/1. DiRAC is
part of the National e-Infrastructure.
We acknowledge the support of the Supercomputing Wales project, which is
part-funded by the European Regional Development Fund (ERDF) via Welsh
Government.
The authors are grateful for computational resources provided by the LIGO
Laboratory, supported by National Science Foundation Grants PHY-0757058 and
PHY-0823459, and by Cardiff University supported by STFC grant ST/I006285/1.

\end{acknowledgments}

\appendix

\section{Neural Network Exploration}
\label{app:nn-explore}

In this appendix we show additional material to justify our choice for the final
networks. As mentioned in Section~\ref{sec:nnts} we run each experiment twice,
first using batched gradient decent and second using mini-batch gradient decent
with a mini-batch size of 1000. We find that the mini-batch results always
outperform the batched results and so we only present the mini-batch results for
most cases.

\subsection{Data Pre-processing:}
\label{app:nn-explore-preproc}

Data pre-processing refers to actions we do to modify the \X and \Y data. We
considered three different options: i) do nothing, ii) \emph{normalise} the data
such that it has zero mean and unit variance or iii) \emph{scale} the data to
lie between 0 and 1. To normalise and scale the data we use the {\tt
StandardScaler} and {\tt MinMaxScaler} functions in the {\tt Scikit-Learn}
python package.

Before performing a more exhaustive search to find the optimal number of hidden
layers and artificial neurons we use an initial network to explore the effects
of data pre-processing. This initial network, found through manual prototyping,
makes use of several common choices in neural network design. It has 6 hidden
layers with 256 neurons in each layer and each neuron uses the rectified linear
unit (ReLU) activation function.

For each dimension or \emph{feature} of \X and \Y we apply the three
pre-processing methods and fit a train a neural network for each pair of
pre-processing methods. We also consider the effect of the mini-batch size on
the training by repeating each experiment twice; once with a batch size equal to
the entire training set ($2 \times 10^{5}$) and again with a mini-batch size
downsampled by 200 giving a mini-batch size of $1000$. We trained the networks
for $10^{3}$ epochs which took $\sim 20$ mins for the batched gradient decent
case and $\sim 40$ mins for the mini-batch case on a Tesla P100 GPU.

We find that the phase \Y data is influenced the strongest by the choice of
pre-processing and the \X data pre-processing has a smaller impact although
is noticeable.
For the amplitude data we find that pre-processing can influence
the results but not as strongly as the phase data.
The reason for this is because the amplitude data is, for the most part,
of the same order of magnitude and O(1). The phase, on the other hand,
as it is accumulated as the binary system evolves can take span
many orders of magnitude depending on the duration of the signal.
Therefore, by applying a pre-processing step such as normalising or scaling
the data brings all the EI coefficients into a similar range which
can help make it easier to train a neural network.
We believe that the pre-processing step applied to the \X data is
less important because, for our dataset, the data is between 0 and 1 for the
$\log(q)$ and between -1 and 1 for the spin dimensions.

We find that, for the phase data, the optimal pre-processing methods are to
normalise the \X data and scale the \Y data. For the amplitude we will
normalise the \X data and use the raw \Y data. We will use these as the
optimal choices for pre-processing the data moving forward and investigate how
the network architecture, choice of optimiser and mini-batch size can effect the
training these neural networks.

\emph{Width and Depth:}
The number of possible configurations a feed-forward, fully-connected artificial
neural network could take presents a near limitless number of possible network
architectures. Whilst the number of neurons in each hidden layer does not have
to be the same we restrict ourselves to neural networks of a constant width
(i.e., number of neurons in each hidden layer) but allows this number and the
number of hidden layers (the depth) to vary. Following the parameterisation in
~\cite{2020arXiv200107523A} perform a systematic search for the optimal number
of hidden layers (depth) $L$ and number of neurons in each hidden layer
(width) $N$. We form the ratio $\beta = L/N$ and consider values $\beta < 1$
which correspond to networks that are wider than their depth. We consider
networks with a maximum number of hidden layers $L_{\rm{max}} = 10$ and three
values of $\beta \in \{0.0125, 0.025, 0.05\}$.

We find that for the phase data $\beta = 0.0125$ predominantly perform best
followed by $\beta = 0.025$ and $\beta = 0.05$ respectively. The same patten is
observed for the amplitude data however, more disordered. The
amplitude data favours deeper networks with $7 - 10$ layers whereas the
phase data prefers networks with $3 - 9$ layers.

For the phase the top two networks both have $\beta = 0.0125$. The best network
has $L = 5$ hidden-layers and $N = 400$ units per layer and the second best
network has $L = 4$ hidden-layers and $N = 320$ units per layer. As the
difference in final loss is insignificant we choose the the network with $4$
hidden-layers as it was significantly faster to train.

For the amplitude data the best performing networks were typically deeper
and wider than the phase networks. However, these differences did not present
a significant increase in accuracy so we opted to use the same network
chosen for the phase data as it also performed well for the amplitude data.

\emph{Activation Function:}
We found that the performance of ANNs on the phase data was strongly influenced
by the choice of activation function but the amplitude data was fairly
insenstive to this choice. For the amplitude the best performing activation
functions were PReLU, ReLU and the Leaky\_ReLU. As the PReLU and the Leaky\_ReLU
adds addition parameters to the training strategy we decided to use the ReLU
activation function for the amplitude. For the phase we find that the PReLU,
ReLU and the Leaky\_ReLU also perform well but the Softplus outperforms them
both in accuracy and training time.

\emph{Optimiser:} We find that SGD, Adadelta and Adagrad consistently
underperform for both the amplitude and phase data, producing loss values $\sim
3$ orders of magnitude worse than the other optimisers tested. For the amplitude
data we find that the Adam optimizer performs equally as well as the Nadam
optimiser and results in a network that is significantly faster to train. For
the phase data we find that AdaMax outperforms Adam, Nadam and RMSprop.

\bibliography{ann}

\end{document}